\pdfoutput=1
\documentclass[aps,prl,twocolumn,superscriptaddress]{revtex4-1}

\usepackage{amssymb,amsmath}
\usepackage{graphics}
\usepackage{dsfont}

\def\nm{{\ \mathrm{nm}}}                       
\def\micron{{\ \mu\text{m}}}                 

\def\Hz{{\ \mathrm{Hz}}}                       
\def\kHz{{\ \mathrm{kHz}}}                     
\def\MHz{{\ \mathrm{MHz}}}                     

\def\us{{\ \mu\mathrm{s}}}                     
\def\ms{{\ \mathrm{ms}}}                       

\def\uT{{\ \mu\text{T}}}                    

\def\uK{{\ \mu\mathrm{K}}}                     

\def\kl{k_L}                            
\def\Rb87{^{87}\mathrm{Rb}}                     

\def\bra#1{\mathinner{\langle{#1}|}}
\def\ket#1{\mathinner{|{#1}\rangle}}

{\catcode`\|=\active
  \gdef\Braket#1{\left<\mathcode`\|"8000\let|\BraVert {#1}\right>}}
\def\BraVert{\egroup\,\mid@vertical\,\bgroup}

\begin{document}


\title{Synthetic partial waves in ultracold atomic collisions}




\author{R.~A.~Williams}
\affiliation{Joint Quantum Institute, National Institute of Standards and Technology, and University of Maryland, Gaithersburg, Maryland, 20899, USA}
\author{L.~J.~LeBlanc}
\affiliation{Joint Quantum Institute, National Institute of Standards and Technology, and University of Maryland, Gaithersburg, Maryland, 20899, USA}
\author{K.~Jim{\'e}nez-Garc{\'i}a}
\affiliation{Joint Quantum Institute, National Institute of Standards and Technology, and University of Maryland, Gaithersburg, Maryland, 20899, USA}
\affiliation{Departamento de F\'{\i}sica, Centro de Investigaci\'{o}n y Estudios Avanzados del Instituto Polit\'{e}cnico Nacional, M\'{e}xico D.F., 07360, M\'{e}xico}
\author{M.~C.~Beeler}
\affiliation{Joint Quantum Institute, National Institute of Standards and Technology, and University of Maryland, Gaithersburg, Maryland, 20899, USA}
\author{A.~R.~Perry}
\affiliation{Joint Quantum Institute, National Institute of Standards and Technology, and University of Maryland, Gaithersburg, Maryland, 20899, USA}
\author{W.~D.~Phillips}
\affiliation{Joint Quantum Institute, National Institute of Standards and Technology, and University of Maryland, Gaithersburg, Maryland, 20899, USA}
\author{I.~B.~Spielman}
\affiliation{Joint Quantum Institute, National Institute of Standards and Technology, and University of Maryland, Gaithersburg, Maryland, 20899, USA}

\date{\today}

\begin{abstract}
Interactions between particles can be strongly altered by their environment.  We demonstrate a technique for modifying interactions between ultracold atoms by dressing the bare atomic states with light,  creating an effective interaction of vastly increased range that scatters states of finite relative angular momentum at collision energies where only $\textit{s}$-wave scattering would normally be expected.  We collided two optically dressed neutral atomic Bose-Einstein condensates with equal, and opposite, momenta and observed that the usual $\textit{s}$-wave distribution of scattered atoms was altered by the appearance of $\textit{d}$- and $\textit{g}$-wave contributions.  This technique is expected to enable quantum simulation of exotic systems, including those predicted to support Majorana fermions.
\end{abstract}


\maketitle


Interactions can be generally understood as arising from collisions between particles which exchange energy and momentum.  For an interacting gas at the lowest temperatures $T$, the thermal de Broglie wavelength $\lambda_{\rm dB}$ can vastly exceed the characteristic range $\ell$ of the interparticle potential; the interactions can then be cast in terms of an effective contact potential that couples states with zero relative angular momentum ($s$-wave interactions).  Only at higher collision energies, when $\lambda_{\rm dB}\lesssim\ell$, do the details of the potential become important.

For alkali atomic gases at degenerate temperatures ($T\lesssim 1\uK$), $\lambda_{\rm dB} \gtrsim 100\nm$ is much larger than the characteristic range of the interatomic van der Waals potential \cite{Chin2010}, $\ell_{\rm vdW} = 1.5\nm$ to $5\nm$, and $s$-wave interactions dominate.  Collisional Fano-Feshbach resonances \cite{Chin2010}, which are used to modify the strength of atomic interactions, do not change the characteristic range of the potential and only \textit{s}-wave interactions are usefully modified \cite{note1}.

Interactions with higher order partial waves are critical for creating systems whose elementary excitations are Majorana fermions, enigmatic particles that are their own anti-particles.  Some many-body systems, including 2D $p$-wave superfluids \cite{Nayak2008} and 1D Fermi gases with spin-orbit coupling \cite{Alicea2010}, have been theoretically shown to support Majorana fermions among their elementary excitations.  Ultracold Fermi gases could realize these systems using laser-induced $p$-wave interactions \cite{Zhang2008,Jiang2011}, the fermionic analog to the bosonic $d$-wave interactions we demonstrate here.

We effectively screen the native atomic interaction with light, reminiscent of effects familiar in condensed matter systems, such as the screening of the Coulomb interaction between a pair of electrons by the collective response of the remaining electrons \cite{Fetter2003}. Similarly, the laser dressing modifies collisions between dressed atoms producing effective higher order partial waves, even though atoms still microscopically collide in the \textit{s}-wave channel.
The interaction potential between a pair of particles located at positions ${\bf r}$ and ${\bf r}^\prime$ can often be described as a function of their separation, $V\left({\bf r} - {\bf r}^\prime\right)$.  The two-body term
\begin{align}
\hat{H}_{\rm int} =& ~\frac{1}{2}\int \frac{d^3{\bf k}_1}{(2\pi)^3}\cdots\frac{d^3{\bf k}_4}{(2\pi)^3} \hat\phi^\dagger({\bf k}_4)\hat\phi^\dagger({\bf k}_3)\nonumber\\
&\times \tilde V({\bf k}_3-{\bf k}_1)\hat\phi({\bf k}_2)\hat\phi({\bf k}_1) \nonumber\\ 
&\times\delta^{(3)}\left({\bf k}_3+{\bf k}_4-{\bf k}_1-{\bf k}_2\right),
\label{Eq1}
\end{align}
in the Hamiltonian describing this interaction is expressed in terms of the Fourier-transformed potential $\tilde V({ \Delta} {\bf k})$, which describes the transition amplitude for a collision that changes the wave vector of a particle by ${ \Delta} {\bf k} = {\bf k}_3-{\bf k}_1$. The field operator $\hat\phi({\bf k})$ denotes the annihilation of a particle with momentum ${\hbar\bf k}=\hbar(k_x,k_y,k_z)$, and the Dirac $\delta$-function ensures conservation of momentum.  The bare interaction between ultracold atoms is well described by a contact potential 
$V({\bf r} - {\bf r}^\prime) \approx g\delta^{(3)}({\bf r}-{\bf r}^\prime)$, with Fourier transform $\tilde V({ \Delta} {\bf k})=g=4\pi\hbar^2 a_s/m$, where $a_s$ is the scattering length and $m$ is the atomic mass.  As $\tilde V({ \Delta} {\bf k})$ is independent of ${ \Delta} {\bf k}$,  low-energy scattering  is isotropic.  The modification of interactions by an environment (screening) can be characterized in terms of a response function \cite{Fetter2003} $\chi$, producing the screened interaction $\tilde V^\prime=\chi({\bf k}_1,{\bf k}_2,{\bf k}_3,{\bf k}_4)\tilde V({\bf k}_3-{\bf k}_1)$.

\begin{figure}[ht]
\begin{center}
\scalebox{0.71}{\includegraphics{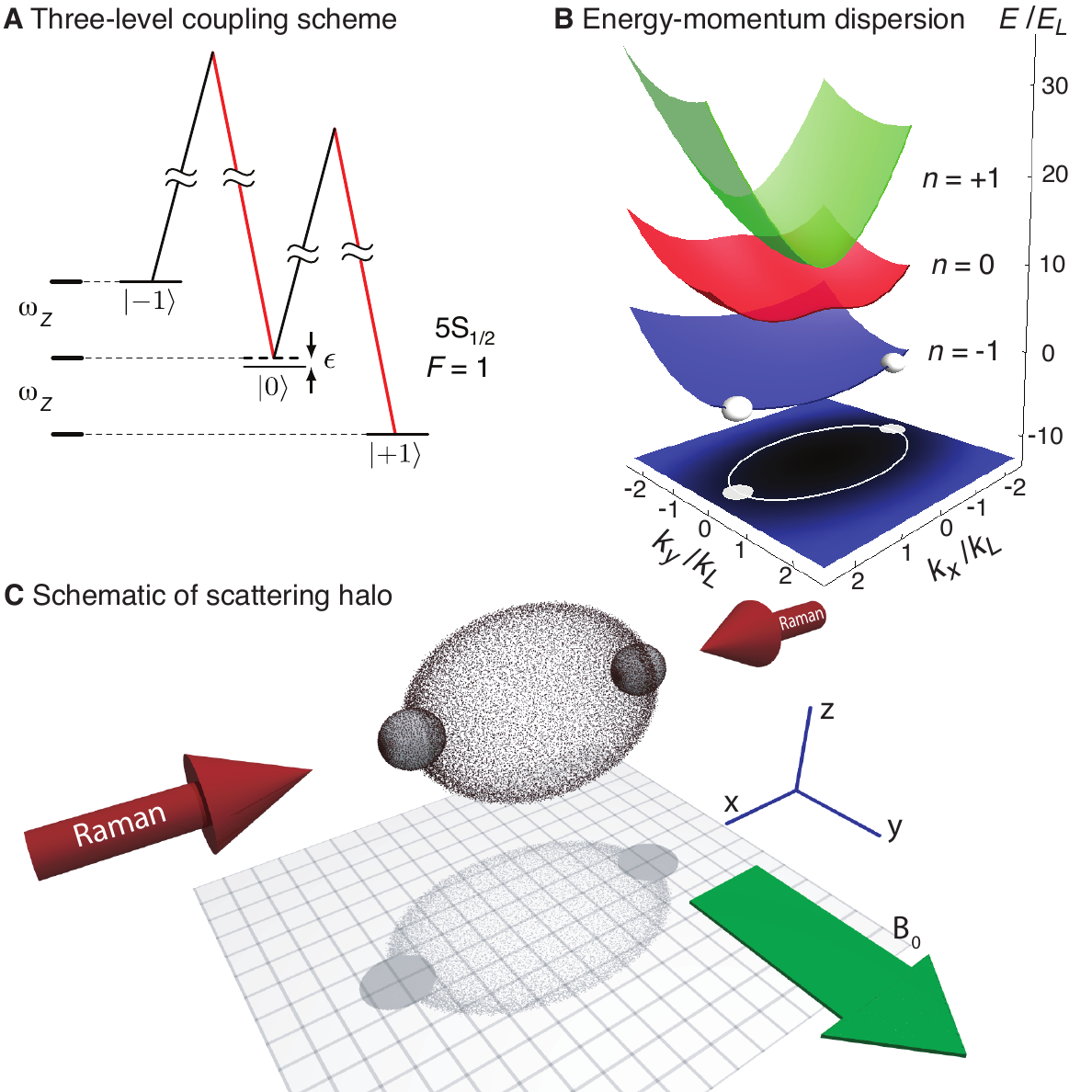}}
\end{center}
\caption{Creation and detection of higher order partial waves. (\textbf{A}) Two Raman laser beams couple the three hyperfine Zeeman states comprising the $F=1$ manifold of the $\Rb87$ $5{\rm S}_{1/2}$ electronic ground state.  (\textbf{B}) Energy-momentum dispersion relations for Raman-dressed atoms with Raman lasers counter-propagating along $\pm {\bf e}_x$.  (\textbf{C}) Schematic representation of two Raman-dressed BECs with atomic momenta $\hbar k_x=\pm 2\hbar k_L$ following their collision, indicating the scattering halo as imaged after TOF.  A magnetic bias field, $B_0{\bf e}_y$, generates the linear ($\omega_z$) and quadratic ($\epsilon$) Zeeman shifts.}
\label{setup}
\end{figure}

To modify the effective atomic interactions we illuminated a Bose-Einstein condensate (BEC) of $\Rb87$ atoms with a two-photon Raman field, producing dressed atoms---spin and momentum superpositions.
Two $\lambda=790.1\nm$ Raman laser beams counterpropagating along $\pm{\bf e}_x$ coupled the three  Zeeman states comprising $\Rb87$'s $F=1$ ground state manifold with strength $\hbar\Omega_R$, (Fig.~1A).  The dispersion relations of laser-dressed atoms (Fig.~1B) are altered from those of free particles, and consist of three distinct bands labeled by $n=\left\{-1,0,+1\right\}$ with energies $E_{0,\pm1}({\bf k})$.  These altered single-particle dispersions lead to synthetic magnetic \cite{Lin2009b} and electric fields \cite{Lin2011a}, and spin-orbit coupling \cite{Lin2011b}.  The current work reveals how Raman-dressing modifies the interaction between atoms.  For atoms in the $n=-1$ (ground) band, we demonstrate higher order effective $d$- and $g$-wave scattering between ultracold atoms at collision energies orders of magnitude below those traditionally required to depart from the $s$-wave regime. Odd partial waves are absent when elastically scattered atoms lie in the same band and are hence indistinguishable, whereas inelastic (band-changing) transitions can have contributions from all partial waves.

We first consider the elastic interactions between dressed atoms in the $n=-1$ band when the relevant energies  (kinetic, interaction, and thermal) are less than the $\approx\hbar\Omega_R$ energy gap between bands.  
The dressed states are related to the bare atomic spin states by a momentum-dependent unitary transformation, $\check U({ \bf k})$ (a $3\!\times\!3$ matrix which depends on $k_x$, not $k_y$ or $k_z$), that links the spin character of an atom to its momentum.
To a very good approximation, interactions between $\Rb87$ atoms in the $F=1$ manifold are spin-independent \cite{Widera2006} and cannot change the spin, hence the scattering amplitude from one pair of momentum states to another depends on the overlap of the spin wavefunctions associated with the initial and final momentum states.  While microscopically the atoms still collide in the \textit{s}-wave channel, the result of this momentum-dependent overlap is that
atoms experience an interaction altered \cite{SOM} by a response function
\begin{align}
\chi =& \bra{n\!=\!-1}\check U({\bf k}_4) \check U^\dagger({\bf k}_2)\ket{n\!=\!-1}\nonumber\\
&\times\bra{n\!=\!-1} \check U({\bf k}_3) \check U^\dagger({\bf k}_1)\ket{n\!=\!-1},
\label{responsefunction}
\end{align}
leading to scattering described by effective higher order partial waves.
The modified scattering can be described by an effective real space potential (determined by the momentum-dependence of $\chi$) that is finite-ranged along the direction of the Raman beams, as discussed in detail in \cite{SOM}.

\begin{figure*}[t]
\scalebox{0.95}{\includegraphics{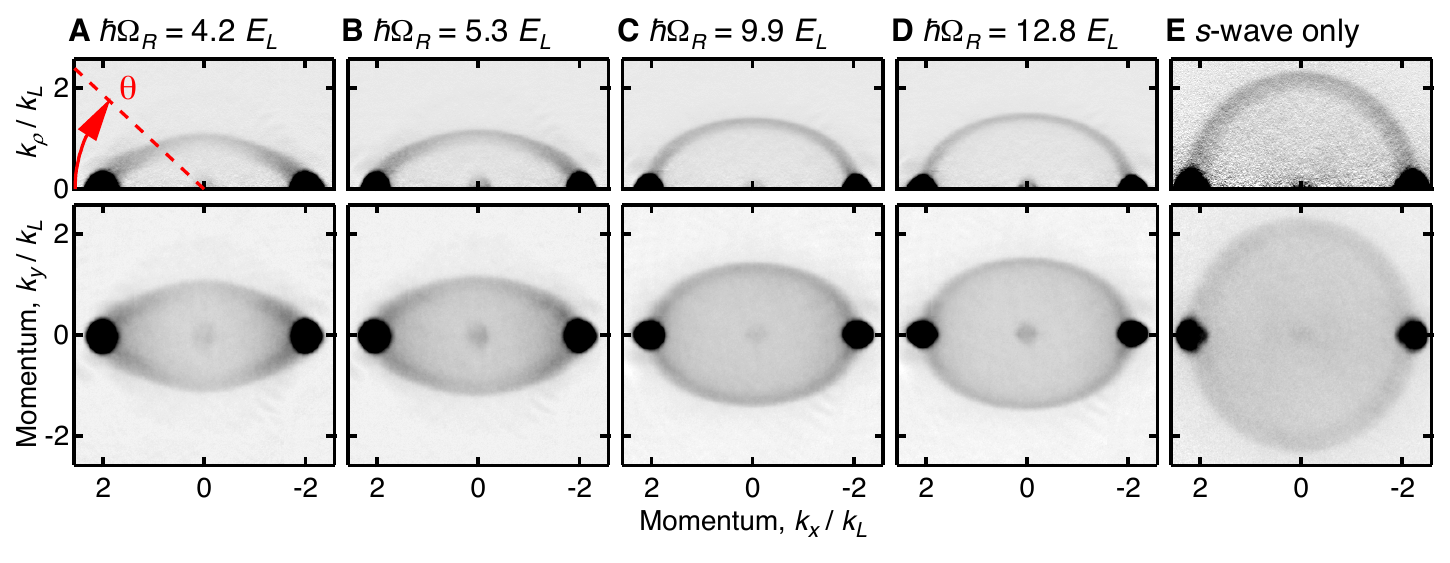}}
\caption{Scattering halos produced by colliding Raman-dressed BECs.  2D momentum distributions, projections of the scattering halo onto the $\mathbf{e}_x$-$\mathbf{e}_y$ plane, are shown along with the radial momentum distributions, $n(k_\rho ,k_x)$, reconstructed using the inverse Abel transform, where $k_\rho=(k_y^2+k_z^2)^{1\slash 2}$.  Darker colors correspond to regions of higher atomic density. (\textbf{A}-\textbf{D}) The halos produced by colliding dressed condensates are nearly ellipsoidal and have a non-isotropic angular density distribution.  For reference, (\textbf{E}) depicts the $s$-wave scattering halo from a pair of colliding BECs without Raman coupling in the $\ket{F=1,m_F=+1}$ spin state.  Each image represents the average of around 50 experimental shots}
\label{ScatteringHaloMontage}
\end{figure*}

We investigated interactions between Raman-dressed atoms by observing the
scattering halo formed by the collision of two BECs in the lowest energy Raman-dressed band (Fig.~1C).
The experiments started with a nearly pure $\Rb87$ BEC of around $5\times 10^5$ atoms in a crossed optical-dipole trap with frequencies, $\{\omega_x,\omega_y,\omega_z\}\slash 2\pi = \{13, 45, 90\}\Hz$.  
That BEC was then prepared in the $n=-1$ band, with coupling strength $\Omega_R$, and split into two (initially spatially overlapping) condensates with momenta $\hbar k_x=\pm 2\hbar k_L$ per atom \cite{SOM}.
After this preparation the dipole trap was immediately turned off ($t_{\rm off}<1\us$) allowing the dressed condensates to collide in the absence of any confining potential.   The scattering halos formed in $\approx 2\ms$ during which time $\Omega_R$ was held constant and the colliding BECs separated and expanded.  After this initial $2\ms$ stage, we transferred the atoms from the ground dressed state into the bare $\ket{F=1,m_F=+1}$ spin state \cite{Lin2011a,SOM}, and imaged the atomic distribution after a $36.2\ms$ time-of-flight (TOF).

Figure~2 shows representative scattering halos from colliding BECs, with (Fig.~2, A-D) and without (Fig.~2E) Raman dressing.  Both the absorption images of the scattering halos (which record the atomic column density in the $\mathbf{e}_x$-$\mathbf{e}_y$ plane) and the radial density distributions $n(k_\rho ,k_x)$, reconstructed using the inverse Abel transform \cite{SOM}, are depicted.
A striking difference between the images in Fig.~2, A-D and that of Fig.~2E is the shape of the scattering halo.  The halo represents a constant energy surface on which the scattered particles lie as a consequence of conservation of energy and momentum in the collision process.  For atoms with an isotropic dispersion relation, the scattering halo is approximately spherical in TOF (Fig.~2E).  However, the Raman lasers break the spherical symmetry of the system by modifying the dispersion relation along ${\bf e}_x$, $E_{-1}(\mathbf{k}) = \mathcal{E}_{-1}(k_x) + \hbar^2(k_y^2 + k_z^2)/2m$, resulting in an approximately ellipsoidal equipotential surface.

As $\Omega_R$ increases, $E_{-1}(\mathbf{k})$ tends to the isotropic $\hbar^2{\bf k}^2/2m$ free-particle dispersion, 
and the scattering halo becomes spherical (seen in the progression in Fig.~2, A-D).  A quantitative analysis of the scattering halos' shape is considered in \cite{SOM}.  We restricted our measurements to $\hbar\Omega_R\gtrsim 4\,E_L$; for smaller coupling strengths the lowest band is significantly anharmonic.  The deviations from sphericity result from a single-particle effect -- the altered dispersion relations -- and are not a signature of beyond $s$-wave scattering.

Along with the modification of scattering halo shape, the distribution of atoms on the scattering surface is also changed.  Two main effects determine the distribution of atoms: (i) the anisotropic final density of states (DOS) $\rho_f({\bf k})$, and (ii) effective higher-order partial waves in the binary collision process.  The first effect, resulting from the shape of the scattering halo as determined by $E_{-1}(\mathbf{k})$, would occur in an identical scattering experiment with spinless bosons subject to the same energy-momentum dispersion relation.  The second effect, described by the response function $\chi$ of Eq.~\ref{responsefunction}, is a consequence of the momentum-dependent spin structure of Raman-dressed atoms leading to effective beyond \textit{s}-wave interactions, and is the central result of this paper.

\begin{figure}[t]
\scalebox{0.8}{\includegraphics{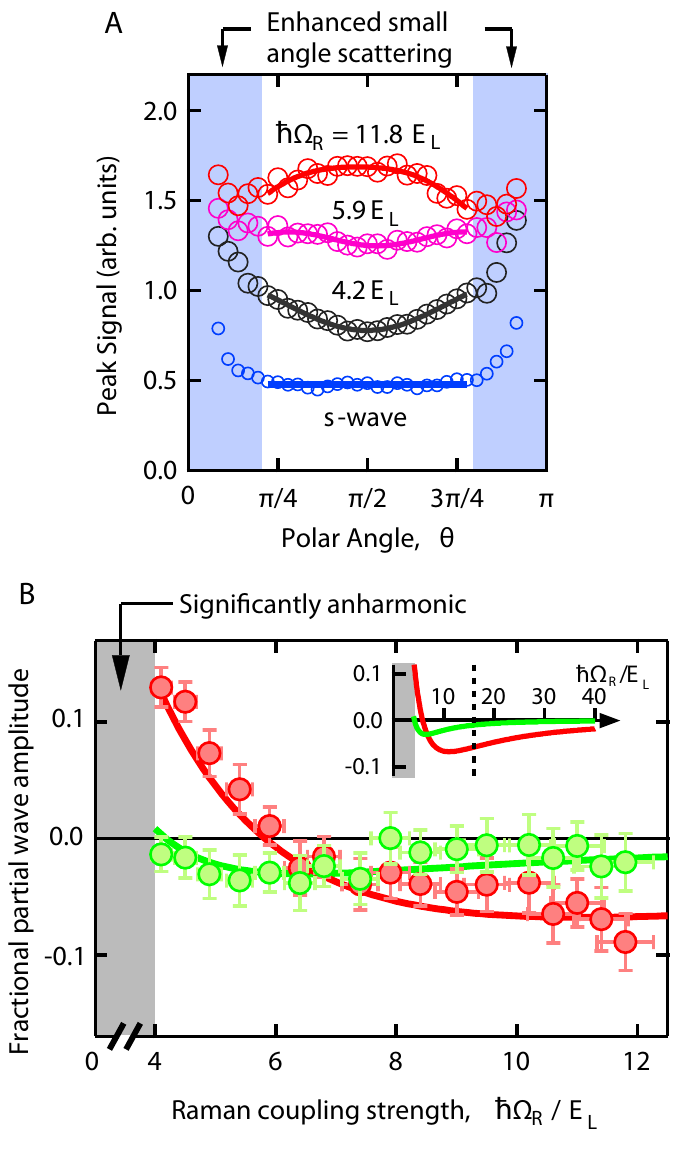}}
\caption{Beyond $s$-wave scattering. (\textbf{A}) Peak signal of the radial momentum distribution (after normalizing to the anisotropic density of states) as a function of $\theta$ (circles) and fits to a partial wave expansion containing \textit{s}-, \textit{d}- and \textit{g}-wave contributions (solid lines).  (\textbf{B}) \textit{d}-wave amplitude (red circles) and \textit{g}-wave amplitude (green circles) as a fraction of the $s$-wave amplitude together with the respective theoretical values (solid lines).  Inset to (\textbf{B}): theoretical fractional partial wave amplitudes plotted over a larger range of $\Omega_R$. For $\hbar\Omega_R / E_L > (\delta k_x /k_L)^2$ (dashed vertical line), the response function $\chi\rightarrow 1$ and the interactions return to pure $s$-wave.}
\label{partialwave}
\end{figure}

These independent effects may be quantified with Fermi's golden rule (FGR).  For two colliding atoms the scattering rate from the initial state $\ket{i}\! =\! {\hat{\phi}^{\prime\dagger}}_{-1}(2 k_L {\bf e}_x){\hat{\phi}^{\prime\dagger}}_{-1}(-2 k_L {\bf e}_x)\ket{{\rm vac}}$ into the final state $\ket{f}\! =\! {\hat{\phi}^{\prime\dagger}}_{-1}({\bf k}_f){\hat{\phi}^{\prime\dagger}}_{-1}(-{\bf k}_f)\ket{{\rm vac}}$ is $\Gamma({\bf k}_f)=$ $2\pi |\langle f|\hat{H}_{\rm int}|i\rangle|^2 \rho_{f}({\bf k}_f)\slash\hbar$, dependent on both the final DOS $\rho_{f}$ and the transition matrix element $\langle f|\hat{H}_{\rm int}|i\rangle$ describing the screened potential $\tilde{V}'$.  Here $\ket{{\rm vac}}$ is the many-body vacuum, and ${\hat{\phi}^\prime}_{n}({\bf k})$ describes the annihilation of a Raman-dressed atom with wave vector ${\bf k}$.  The dependence of the matrix element on the response function $\chi$ gives rise to effective higher order partial wave scattering \cite{SOM}.  The contributions of the DOS and the matrix element to the anisotropic nature of the scattering rate $\Gamma({\bf k}_f)$ are of comparable magnitude.

Figure~3A shows the peak signal as a function of polar angle $\theta$ extracted from the inverse Abel transform
for several different $\Omega_R$, normalized to the anisotropic DOS \cite{SOM}.  This normalization removes all anisotropies except those resulting from beyond $s$-wave scattering, which we quantify by fitting the data with the partial-wave expansion \cite{Sakurai1993},
$p(\theta) = A|\sum_{l=0,2,4}{(2l+1)(\exp{2 i\eta_l}-1)P_l(\cos{\theta})}|^2\slash k^2$,
where the fit parameter, $\eta_l$, is the phase shift of the $l^\mathrm{th}$ partial wave, $P_l(\cos{\theta})$ is the Legendre polynomial of order $l$, and $A$ is an overall scaling factor (e.g., dependent on the initial condensate densities and their interaction time).  The peak signal $p(\theta)$ is proportional to $\Gamma({\bf k}_f)$.  The fit was restricted to the angular range $\pi\slash 5\leq\theta\leq 4\pi\slash 5$ which excluded the regions close to the unscattered clouds. These regions showed an increase in atom number beyond that expected, even for pure \textit{s}-wave scattering between bare spin states, possibly because of multiple scattering events or Bose stimulation.

Figure~3B displays the \textit{d}- and \textit{g}-wave amplitudes (normalized to the \textit{s}-wave amplitude) as a function of $\Omega_R$, obtained from fits as in Fig.~3A, where the fractional amplitude of the $l^{{\rm th}}$ partial wave is $(2l+1) {\rm sign}(\eta_0\eta_l)\left|\exp{2 i\eta_l}-1\right|\slash\left|\exp{2 i\eta_0}-1\right|$.  At $\hbar\Omega_R=4.1 E_L$, we find a fractional $d$-wave amplitude of $13\%$ for collision temperature $T_{\rm col}=\mu v^2_{\rm rel}\slash 2 k_B\approx 0.7\uK$, where $\mu$ is the reduced mass of the colliding particles and $v_{\rm rel}$ their relative velocity.  By comparison, two previous experiments \cite{Thomas2004,Buggle2004}, which studied the scattering of $\Rb87$ in a bare spin state, did not find comparable $d$-wave magnitudes until $T_{\rm col}\gtrsim 100\uK$.   The solid lines show the theoretical prediction for the fractional partial wave amplitudes, calculated from a partial wave expansion of the matrix element $\langle f|\hat{H}_{\rm int}|i\rangle$.
Figure~3B highlights that the engineered effective interaction potentials are not only long-ranged but also tunable;  by changing $\Omega_R$ we controlled both the magnitude and sign of the $d$-wave phase shift.

We now consider collisions between atoms in the highest energy $n=1$ Raman-dressed band.
Although spin-changing collisions of bare atoms hardly occur in $\Rb87$, strong band-changing transitions of dressed atoms are in general permitted \cite{StamperKurn2003,Spielman2009}.
To study this effect we investigated the (energetically allowed) collision-induced transitions of atoms to lower bands. 
We considered collisions in a single BEC (so atoms collided at extremely low relative momenta) and
observed the spatial distribution of the collision products after TOF.  

We prepared a BEC in the highest energy dressed state with zero momentum \cite{SOM}.
Figure~4B shows atoms in the $n=0$ (intermediate energy) band which have decayed from the excited band by the processes indicated in Fig.~4A.  Scattered atoms exhibited a node in the $k_x=0$ plane, a dramatic manifestation of beyond \textit{s}-wave scattering.  The inner halo (I) is due to a process where one atom in a colliding pair undergoes an inelastic transition to the $n=0$ band while the other scattered atom remains in the $n=1$ band.  As the scattering products are distinguished by their band index, the spatial wave function need not be symmetric with respect to exchange, and the scattered wave function can in principle have contributions from all partial waves.  However, only effective $p$-wave and higher odd partial wave components are permitted due to the underlying symmetry of the Hamiltonian.

\begin{figure}[t]
\begin{center}
\scalebox{0.8}{\includegraphics{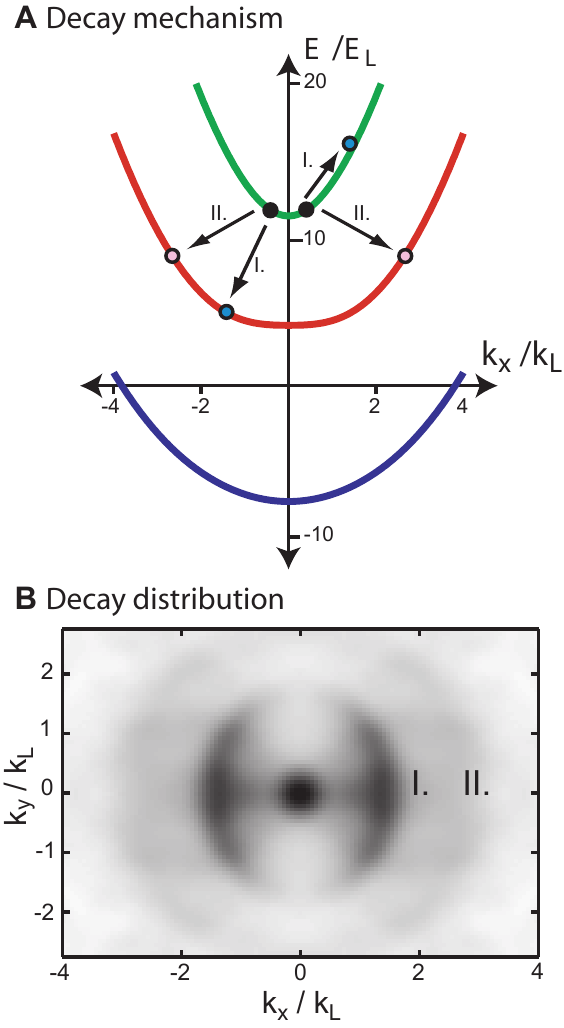}}
\end{center}
\caption{Collision-induced decay from the $n=1$ excited Raman-dressed state. (\textbf{A}) Decay paths from the $n=1$ excited state to the $n=0$ state (here $\hbar\Omega_R=13\,E_L$). (\textbf{B}) Momentum distribution of decay products in the $n=0$ state.  After loading into the $n=1$ band the atoms were held in the dipole trap for $1\ms$ before release, allowing a significant fraction to decay to the lower energy bands.  An adiabatic deloading technique transferred $n=0$ dressed atoms into the $m_F=0$ spin state \cite{SOM}, and the atoms were subsequently imaged after $7\ms$ TOF.
}
\label{excitedstate}
\end{figure}

Our technique for modifying atomic interactions has significant ramifications for systems of ultracold atoms, bosons and fermions alike.  Possible new effects induced by synthetic higher-order partial wave interactions in a harmonically trapped BEC could include roton features in the excitation spectrum \cite{StamperKurn2003}, crystalline ordering \cite{Romanovsky2004}, supersolidity \cite{Scarola2005}, or even the support of 2D solitons \cite{Tikhonenkov2008}.  In addition, collisions in this experiment are subject to a new kind of resonance phenomena.  As the collision energy approaches the inter-band spacing, virtual excitations to the higher band can greatly modify the collisional properties beyond the effects described here.

\begin{acknowledgments}
We thank C. Sa de Melo, J.~V.~Porto, and V.~Galitski for useful conversations.  This work was partially supported by ONR; by ARO with funds from the DARPA OLE program and the Atomtronics MURI; and by the NSF through the Physics Frontier Center at JQI.  L.~J.~L. acknowledges NSERC, K.~J.-G. acknowledges CONACYT, M.~C.~B. acknowledges NIST-ARRA.
\end{acknowledgments}


\section*{Methods and Materials}
\subsection*{Dressed states}
Consider the operator $\hat{\varphi}_\sigma({\bf { q}})$ which destroys a particle  with momentum ${\hbar\bf { q}}$ and $\sigma = 0,\pm 1$, where $\sigma$ denotes $m_F = 0,\pm 1$.
To describe the states coupled by the Raman dressing it is convenient to introduce the operators $\{ \hat{\phi}_{-1}(\tilde{\bf k}),\, \hat{\phi}_{0}(\tilde{\bf k}),\, \hat{\phi}_{+1}(\tilde{\bf k})\} =\{ \hat{\varphi}_{-1}(\tilde{\bf k}+2k_L{\bf e}_x),\,  \hat{\varphi}_{0}(\tilde{\bf k}),\, \hat{\varphi}_{+1}(\tilde{\bf k}-2k_L{\bf e}_x)\}$, where the quantum number $\hbar\tilde{\bf k}$ is sometimes described as the "quasimomentum," in analogy to the appropriate quantum number for dressed (Bloch) states in a lattice potential.  From here on and throughout the paper we drop the tilde notation and refer to $\hbar{\bf k}$ as the "momentum" of a dressed particle.
Our experimental system (Fig.~1) is described by the Hamiltonian
\setcounter{equation}{0}
\renewcommand{\theequation}{S\arabic{equation}}
\begin{equation}
\hat{H} = \int{\frac{d^3{\bf k}}{(2\pi)^3}\sum_{\sigma,\sigma'}{\hat{\phi}^\dagger_\sigma({\bf k}) H_{\sigma,\sigma'}({\bf k})\hat{\phi}_{\sigma'}({\bf k})}}  + \hat{H}_{\rm int},
\end{equation}
where $H_{{\sigma,\sigma'}}$ is an element of the $3\times 3$ matrix  $\check{H} = \check{H}_1(k_x) + [\hbar^2(k_y^2+k_z^2)\slash 2m]\otimes\check{\bf 1}$ and $\hat{H}_{\rm int}$ describes binary interactions between the particles (discussed below). The free motion along ${\bf e}_x$, together with Zeeman shifts and atom-light coupling, is described by $\check{H}_1(k_x)$, $\check{\bf 1}$ is the $3\times 3$ identity matrix.  (We do not consider an external potential as the condensates collided after the optical dipole trap had been turned off and the scalar light shift induced by the Raman beams was made negligible by operating at the magic wavelength $\lambda = 790.1\nm$).
A static bias magnetic field $B_0\,{\bf e}_y$ with $B_0 \approx 406 \uT$ generated a $\omega_{\rm Z}\slash 2\pi = 3.25\MHz$ Zeeman shift between the bare spin states $m_F = 0,\pm 1$, with a small $\hbar\epsilon = 0.42\,E_L$ offset of the $m_F=0$ state due to the quadratic part of the Zeeman shift ($E_L\slash h = 3.6\kHz$). The counterpropagating Raman beams had orthogonal linear polarizations and a frequency difference $\Delta\omega_L$, (giving a detuning $\delta = \Delta\omega_L-\omega_z$ from resonance) coupling the bare states $\ket{m_F=-1,k_x + 2k_L}$, $\ket{m_F=0,k_x}$, $\ket{m_F=+1,k_x - 2k_L}$ with strength $\Omega_R$.
In this bare basis and in the rotating frame $\check{H}_1(k_x)\slash\hbar$ has the form \cite{Lin2009a}
\begin{align*}
\begin{pmatrix}
\frac{\hbar}{2m}(k_x + 2k_L)^2 - \delta & \Omega_R\slash 2 & 0\\
\Omega_R\slash 2  & \frac{\hbar}{2m}(k_x)^2-\epsilon & \Omega_R\slash 2 \\
0  & \Omega_R\slash 2  & \frac{\hbar}{2m}(k_x - 2k_L)^2 + \delta
\end{pmatrix}.
\label{eq:matrixHamiltonian}
\end{align*}

For the experiments described in this paper $\delta=0$ except for a final deloading step (discussed below) after the condensates had collided (the deloading procedure mapped dressed eigenstates, which are spin-momentum superpositions, back to a single bare spin-momentum state).
For a given value of $k_x$, $\check{H}_1$ can be diagonalized by a $3\times 3$ unitary matrix, $\check{U}({\bf k})$, and the Raman-dressed eigenstates are related to the bare states by the momentum-dependent unitary transformation:
$\hat{\phi}'_n({\bf k}) = \sum_{\sigma}{ U_{n,\sigma}({\bf k})\hat{\phi}_{\sigma}({\bf k})}$, where $n$ is the band-index of the dressed states.


\subsection*{System preparation}
Our experiments began with nearly pure $\Rb87$ BECs of around $5\!\times\! 10^5$ atoms in the $\ket{F=1,m_F=-1}$ state confined in a crossed optical dipole trap.  The trap consisted of a
pair of $1064\nm$ laser beams propagating along ${\bf e}_x$
($1/e^2$ radii of $w_{y}\!\approx\! 120\micron$ and $w_{z}\!\approx\! 50\micron$)
and ${\bf e}_y$ ($1/e^2$ radii of $w_{x}\!\approx\! w_{z}\!\approx\! 160\micron$).
The trap frequencies at the end of an evaporation sequence were
$\{\omega_x,\omega_y,\omega_z\}/ 2\pi = \{13, 45, 90\}\Hz$.

To study collisions beyond $s$-wave we prepared colliding Bose-Einstein condensates in the lowest energy Raman-dressed band using the following procedure:
(i) starting with an $\ket{F=1,m_F=-1}$ BEC at rest, a radio-frequency magnetic field, $B_{\text{rf}}(t) {\bf e}_x$,
was ramped on in $100\ms$ with an initial detuning $\hbar\delta_{\text{rf}} = -26\,E_L$ (where $\delta_{\text{rf}}=\omega_{\text{rf}}-\omega_Z$, and the detuning $\delta_{\text{rf}}$ was controlled by changing $B_0$) and coupling strength $\hbar\Omega_{\text{rf}}= 3\,E_L$; (ii) the detuning was adiabatically ramped
to $\delta_{\text{rf}} \approx0$ in $25\ms$, and then the radio-frequency coupling strength $\Omega_{\text{rf}}$ was decreased to
zero in $12\ms$.  In the last millisecond of the ramp of $\Omega_{\text{rf}}$ the dressed state to which we connect adiabatically is a virtually equal superposition of $m_F=\pm1$; (iii) During this final millisecond $\Omega_R$ increased from zero to its final
value, creating two spatially overlapping condensates in the ground dressed band with
momenta $\pm 2\hbar k_L$.  The turn-on of the Raman beams was adiabatic with respect to inter-band excitation but fast with respect to the trap's $\approx20\ms$ quarter-period along ${\bf e}_x$.  This ensured the atoms could not change their momentum as $\Omega_R$ was increased \cite{Lin2009a}.

To observe $s$-wave scattering we prepared the system in the absence of any Raman beams.
The BEC,
initially formed in the $\ket{F=1,m_F=-1}$ state, was transferred to the $\ket{F=1,m_F=+1}$ state by radio-frequency adiabatic rapid passage. Thus the atoms were in the same spin state as for the time-of-flight (TOF) stage for the higher partial wave data (see Deloading below), ensuring that
any small distortions of the scattering halo in TOF expansion due to stray magnetic field gradients
were the same for the $s$-wave data and the higher order partial wave data.
Two temporally square pulses of a 1D optical lattice along ${\bf e}_x$ (formed with $\lambda = 792.1\nm$ light) were used to split the BEC,
initially at rest, into $\pm 2\hbar k_L$ orders \cite{Wu2005}.  The dipole trap
was switched off ($t_{\rm off}<1\us$) at the end of the second optical lattice pulse, and the
resulting atomic distribution absorption-imaged after $36.2\ms$ TOF. 

To prepare a BEC in the highest energy dressed state we first loaded the BEC into the lowest energy Raman dressed state
with zero momentum \cite{Lin2011a, Lin2009a}. The BEC was then excited to the highest energy dressed state with zero momentum by amplitude modulating the Raman beams.  For a coupling strength $\Omega_R=13\,E_L$, the amplitude of the Raman beams was modulated by $17\%$ (half peak-to-peak amplitude) for $1\ms$.  During this $1\ms$ the frequency of the modulation was swept across the resonant energy difference $\Delta E$ between the ground and excited dressed states at zero
momentum, from $\Delta E\slash\hbar-3\kHz$ to $\Delta E\slash\hbar+3\kHz$, allowing near complete population transfer to the highest energy dressed state by adiabatic rapid passage ($\Delta E = 18.9\,E_L = h\times 69.4\kHz$ for $\Omega_R=13\,E_L$).  The $1\ms$ sweep time represented a compromise between being slow enough for adiabatic transfer while being fast enough to avoid significant decay from the excited state during the transfer.

\subsection*{Inverse Abel transform}
We used the inverse Abel transform \cite{Bracewell2000} to reconstruct the radial density distribution, $n(\rho,z)$, from the column densities $n(x,y)$ recorded in our absorption images.
The inverse Abel transform is appropriate for axially symmetric distributions, which is the case for the scattering products after TOF.  The small central cloud of atoms in the images (Fig.~2) is due to a residual population of $k=0$ atoms remaining from the loading process into the ground Raman-dressed band.  The density of the $k=0$ clouds is small enough that their effect on the scattering halos is negligible.

\subsection*{Deloading}
Raman-dressed atoms are in a superposition of spin and momentum states.  When the Raman beams were turned off diabatically, the Raman-dressed atoms were projected onto a superposition of bare spin states $\ket{m_F\!=-1,k_x+2k_L}$, $\ket{m_F\!=0,k_x}$, $\ket{m_F\!=+1,k_x-2k_L}$, somewhat complicating the atomic distribution in TOF.  In order to image a single, well-formed scattering halo, we adiabatically mapped atoms in the $n\in\left\{-1,0,+1\right\}$ band to the $\ket{m_F = -n, k_x+2k_L n}$ bare state.  This was achieved by ramping the detuning of the Raman beams from $\hbar\delta=0$ to $\hbar\delta = 50\,E_L$ (for which the Raman-dressed states coincide nearly perfectly with pure spin states).  The $2\ms$ long ramp was started after $2\ms$ of TOF, by which time the colliding condensates had spatially separated.  After the ramp $\Omega_R$ was turned off in $100\us$, leaving the atoms to continue their time-of-flight in the $m_F=+1$ state.  In analyzing the data we neglect the change in group velocity during the few milliseconds TOF during which the atoms are dressed.  This has a negligible effect.

\subsection*{Effective higher order partial waves}
We now consider the term describing binary interactions between particles, $\hat{H}_{\rm int}$.  The four-field interaction term for spin-independent (a very good approximation for $\Rb87$ atoms \cite{Widera2006} but not necessarily for other atomic species) contact interactions of the form $V({\bf r} - {\bf r}^\prime) = g\delta({\bf r}-{\bf r}^\prime)$ is
\begin{align}
\hat{H}_{\rm int} =& \,\frac{g}{2} \int{d^3{\bf r}\,\sum_{\sigma_1,\sigma_2}{\hat{\psi}^\dagger_{\sigma_1}({\bf r})\hat{\psi}^\dagger_{\sigma_2}({\bf r})\hat{\psi}_{\sigma_1}({\bf r})\hat{\psi}_{\sigma_2}({\bf r})}}\\ \nonumber
=& \int \frac{d^3{\bf k}_1}{(2\pi)^3}\frac{d^3{\bf k}_2}{(2\pi)^3}\frac{d^3{\bf k}_3}{(2\pi)^3}\frac{d^3{\bf k}_4}{(2\pi)^3}\hat{\tilde V}({\bf k}_1,{\bf k}_2,{\bf k}_3,{\bf k}_4),\nonumber
\end{align}
with
\begin{align}
\hat{\tilde  V}({\bf k}_1,{\bf k}_2,{\bf k}_3,{\bf k}_4)=& \frac{g}{2}\sum_{\sigma_1,\sigma_2}
\hat\phi_{\sigma_1}^\dagger({\bf k}_4)\hat\phi_{\sigma_2}^\dagger({\bf k}_3)\hat\phi_{\sigma_1}({\bf k}_2)\hat\phi_{\sigma_2}({\bf k}_1)\nonumber\\ 
&\times \,\delta^{(3)}\left({\bf k}_3+{\bf k}_4-{\bf k}_1-{\bf k}_2\right).
\end{align}
Eq.~S4 shows that the matrix element for a collision between the bare states is independent of the momenta of the colliding particles.

Transforming to the dressed state basis the interaction Hamiltonian becomes
\begin{align}
\label{Vtildehat}
\hat{\tilde  V} =& \,\frac{g}{2}\sum_{n_1',n_2',n_1,n_2}
\hat\phi{'}_{n_1'}^\dagger({\bf k}_4)\hat\phi{'}_{n_2'}^\dagger({\bf k}_3)\hat\phi'_{n_1}({\bf k}_2)\hat\phi'_{n_2}({\bf k}_1)\nonumber\\ 
&\times\left[ \sum_{\sigma_1,\sigma_2}{U_{n_1',\sigma_1}({\bf k}_4)U^\dagger_{\sigma_1,n_1}({\bf k}_2)U_{n_2',\sigma_2}({\bf k}_3)U^\dagger_{\sigma_2,n_2}({\bf k}_1)}\right]\nonumber\\ 
&\times \delta^{(3)}\left({\bf k}_3+{\bf k}_4-{\bf k}_1-{\bf k}_2\right).
\end{align}
The key point is that the transition amplitude for a process in which dressed particles with initial wave vectors ${\bf k}_1$ and ${\bf k}_2$ are scattered to final wave vectors ${\bf k}_3$ and ${\bf k}_4$ is now dependent on ${\bf k}_1,{\bf k}_2,{\bf k}_3,{\bf k}_4$ through the elements of the unitary matrices.  For example if the initial and final scattering states are constrained to lie within the lowest band, we have $\ket{i} = \hat{\phi}{'}_{-1}^\dagger({\bf k}_1)\hat{\phi}{'}_{-1}^\dagger({\bf k}_2)\ket{{\rm vac}}$, $\ket{f} = \hat{\phi}{'}_{-1}^\dagger({\bf k}_3)\hat{\phi}{'}_{-1}^\dagger({\bf k}_4)\ket{{\rm vac}}$, and the matrix element between these states is
\begin{align}
\langle f|\hat{H}_{\rm int}|i\rangle
=& \,\frac{g}{2}\big[\chi({\bf k}_1,{\bf k}_2,{\bf k}_3,{\bf k}_4)+\chi({\bf k}_2,{\bf k}_1,{\bf k}_3,{\bf k}_4)\nonumber\\ 
& +\chi({\bf k}_1,{\bf k}_2,{\bf k}_4,{\bf k}_3)+\chi({\bf k}_2,{\bf k}_1,{\bf k}_4,{\bf k}_3)\big]\nonumber\\ 
\times & \delta^{(3)}\left({\bf k}_3+{\bf k}_4-{\bf k}_1-{\bf k}_2\right),
\end{align}
where
\begin{align}
\chi({\bf k}_1,{\bf k}_2,{\bf k}_3,{\bf k}_4) &= \sum_{\sigma_1,\sigma_2}U_{-1,\sigma_1}({\bf k}_4)U^\dagger_{\sigma_1,-1}({\bf k}_2)\nonumber\\
&\times U_{-1,\sigma_2}({\bf k}_3)U^\dagger_{\sigma_2,-1}({\bf k}_1).
\label{eq:chiExpression}
\end{align}
Moreover, it is apparent from Eq.~\ref{Vtildehat} that the matrix elements for band-changing collision events are in general non-zero, and also dependent on the initial and final momenta of the colliding particles.  In the lowest energy Raman-dressed band such band-changing collisions are energetically forbidden provided the collision energy is small compared to the gap to the excited bands (as is the case in this paper).  For particles in the excited bands collision-induced decay to lower energy bands is possible as discussed in the main body of the paper and shown in Fig.~4.

The total elastic scattering cross section is given by the sum of cross sections of each partial wave contribution, however the total collision cross section does not increase when the atomic interactions are modified by the laser dressing.  This is because rather than adding $d$-wave and $g$-wave contributions on top of the $s$-wave scattering component, the interactions are modified such that part of the $s$-wave contribution is renormalized to $d$- and $g$-wave contributions.  The total collision cross section is in fact generally smaller than the bare state collision cross section, due to the partial (spin-dependent) overlap between the initial and final scattering states.  The modified density of states in the presence of laser dressing also reduces the number of final states for a particle to scatter into.

\setcounter{figure}{0}
\renewcommand{\thefigure}{S\arabic{figure}}
\begin{figure}[t]
\scalebox{0.90}{\includegraphics{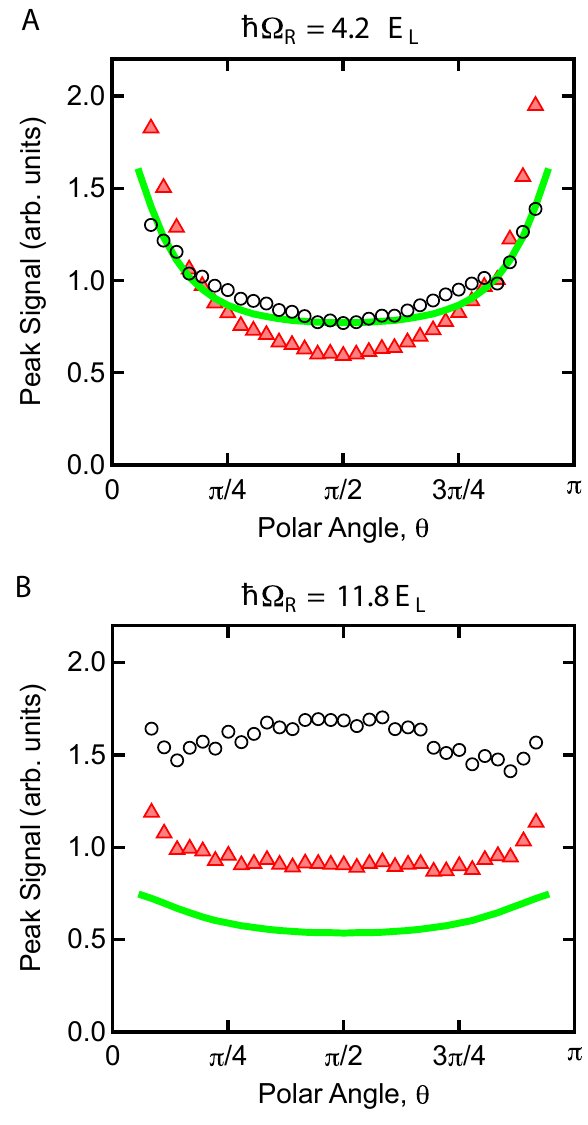}}
\caption{Contributions of density of states (solid green line) and higher order partial wave scattering (black circles) to the $\theta$ dependence of the measured scattered atomic distribution (red triangles). The higher order partial wave scattering signal is given by the ratio of the red triangles to the solid green line (\textbf{A}) $\hbar\Omega_R = 4.2~E_L$. (\textbf{B}) $\hbar\Omega_R = 11.8~E_L$.}
\label{DOS_graph}
\end{figure}
\subsection*{Density of states}  The density of states on the scattering halo is purely geometric in origin and depends on the coupling strength $\Omega_R$, which determines the shape of the scattering halo through the modified dispersion relation $E({\bf k})$.  Since $\Omega_R$ is measured (see below) for each scattering halo, one can calculate the form of the density of states and remove its effect from the measured atomic signal, revealing the underlying anisotropic atomic distribution due to the modification of the interaction potential by the response function $\chi({\bf k}_1,{\bf k}_2,{\bf k}_3,{\bf k}_4)$.  The uncertainty introduced by dividing out the density of states is set by the uncertainty in $\Omega_R$, from which the bandstructure and hence the density of states are calculated.  The Raman coupling strength was measured using a Raman pulsing technique \cite{Lin2009a}.  The uncertainty (standard deviation) in $\Omega_R$ from this measurement was $\pm 5\,\%$, translating to a comparable or smaller uncertainty in the extracted partial wave amplitude (determined by carrying out the partial wave fitting procedure on data which had been normalized to the density of states calculated for $\Omega_R\pm5\,\%$).  The uncertainties in the extracted partial-wave amplitudes of Fig.~3\,B represent the combined uncertainties from the partial wave fits to the data (which dominate) and the uncertainty in the density of states.

Fermi's golden rule (FGR) states that the transition rate from an initial state $\ket{i}$ to a final state $\ket{f}$ due to a coupling $\hat{H}_{\rm int}$ is $\Gamma_{if} = 2\pi|\langle f|\hat{H}_{\rm int}|{i}\rangle|^2 \,\delta\left(E_i-E_f\right)\slash\hbar$.  If one presumes there is some parameter, i.e.\ the wave vector ${\bf k}$, over which the final states are uniformly distributed then the total transition rate can be written
\begin{equation}
\Gamma_i = \frac{2\pi}{\hbar} \int{\frac{{\rm d}^3{\bf k}}{(2\pi)^3}|\langle f|\hat{H}_{\rm int}|{i}\rangle|^2}\,\delta\left(E_i-E({\bf k})\right).
\end{equation}
Considering a function ${\bf K}(\theta)$ that gives the momentum as a function of $\theta$ on the energy surface $E({\bf K})$, the total scattering rate can be written
\begin{equation*}
\Gamma_i = \frac{1}{\hbar}\int_0^\pi{[K(\theta)]^2\sqrt{1+\frac{[K'(\theta)]^2}{[K(\theta)]^2}}\sin{\theta}{\rm d}\theta \left[\frac{|\langle i|\hat{H}_{\rm int}|{\bf K}\rangle|^2}{\nabla_k E({\bf K})}\right]}.
\end{equation*}
where $K^\prime$ is the derivative of $K$ with respect to $\theta$. 
The measured peak atomic signal at a given angle, $p(\theta)$, is proportional to $\Gamma$, and hence proportional to the quantity in square brackets.  Figure~\ref{DOS_graph} displays the peak atomic signal as a function of polar angle $\theta$ (red triangles), as directly measured from the radial momentum distribution, as well as the calculated contribution from the density of states (solid green line).  The ratio of these is the contribution attributed to beyond $s$-wave scattering (black circles). It is this latter signal which is shown in Fig.~3 A.

\subsection*{Shape of the scattering halos}
\begin{figure}[t]
\begin{center}
\scalebox{0.9}{\includegraphics{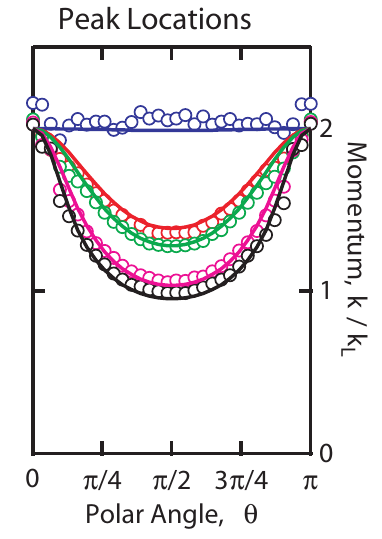}}
\end{center}
\caption{Experimentally determined location (circles) of the scattering peaks, $k=\left(k_\rho^2+k_x^2\right)^{1/2}$ in Fig.~2, A-D (see main text) as a function of polar angle $\theta$, together with theory (solid lines).  The theory curves have no adjustable parameters and neglect mean-field-driven expansion.  Black - $4.2\,E_L$; magenta - $5.3\,E_L$; green - $9.9\,E_L$; red - $12.8\,E_L$; blue - $s$-wave.   The $s$-wave halos displayed small deviations from a perfectly spherical form, as expected for an initially anisotropic condensate for which mean-field interactions lead to a detectable distortion of the halo at the few percent level \cite{Krachmalnicoff2010}.}
\label{PeakLocation}
\end{figure}

As $\Omega_R$ increases, $E_{-1}(\mathbf{k})$ tends to the isotropic $\hbar^2{\bf k}^2/2m$ free-particle dispersion, 
and the scattering halo becomes spherical, as can be seen in the progression in Fig.~2, A-D (main text).  Figure~\ref{PeakLocation} quantifies the shape of the scattering halos, by plotting the positions of the peak in $n(k_\rho ,k_x)$ as a function of the polar angle $\theta$.  The peak locations are in agreement with those expected from numerical calculations of $E_{-1}({\bf k})$ using the measured values of $\Omega_R$.

\subsection*{Lifetime in the $n=+1$ excited Raman band}
We observed a $\tau=12(1)\ms$ $1/e$ lifetime for decay from the excited state, in reasonable agreement with an approximate calculation of the initial decay rate due to binary collisions in our BEC based on FGR \cite{Spielman2006}, predicting a lifetime $\tau\approx 9\ms$.  These excited bands are of intrinsic interest, for example in schemes considering oppositely-charged particles in synthetic gauge fields or proposals for generating non-abelian gauge fields for ultracold atoms, which rely on two excited dressed states (see \cite{dalibard2010} and references therein).

\begin{figure}[t]
\scalebox{0.9}{\includegraphics{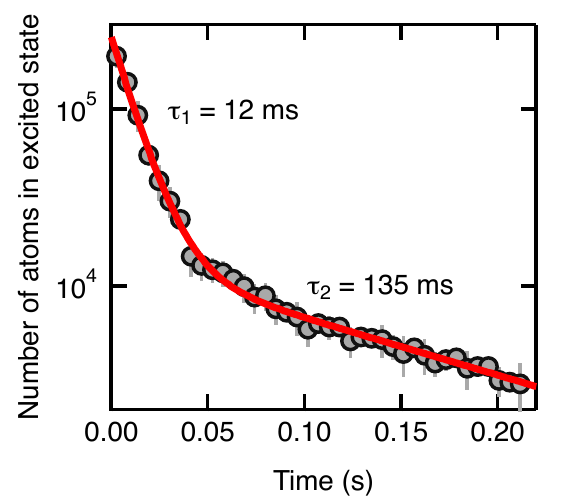}}
\caption{Number of atoms in the excited Raman-dressed state as a function of hold time in the optical dipole trap (circles).  The solid red line is a fit to a double exponential function.}
\label{ExcitedStateDecayLifetime}
\end{figure}

In order to measure the lifetime in the $n=+1$ band we excited a BEC in the $|n=-1,{\bf k}=0\rangle$ state to the $|n=+1,{\bf k}=0\rangle$ state as described above.  The strength of the Raman coupling for this lifetime experiment was $\Omega_R = 7~E_L$. 
Figure~\ref{ExcitedStateDecayLifetime} shows the number of atoms in the excited state as a function of hold time in the optical dipole trap.
For simplicity a double exponential fit of the form $N_{n=+1}(t)= A_1\exp(-t\slash\tau_1)+A_2\exp(-t\slash\tau_2)$ was used to the extract the initial decay time of $\tau_1 = 12\ms$.  We attribute the slower decay, $\tau_2 = 135\ms$, to the depletion of excited state BEC and subsequent decay of the lower density thermal atoms.
 Note that in the limit of very large Raman coupling, $\hbar\Omega_R\gg E_L$, (which was outside our range of available laser powers) the matrix element for these band-changing collisions tends to zero, and long lifetimes in the excited state should be possible.  It would also be possible to suppress collisional losses from the excited state at lower $\Omega_R$ by modifying the final density of states, for example, by the use of an optical lattice to open band gaps at the relevant energies.

\subsection*{Real space representation of the effective interaction potential}
That the response function $\chi$ depends on the initial and final momenta of the colliding particles implies that an effective real space potential describing the screened collisions can no longer be represented by a Dirac $\delta$-function but has some finite range.
Furthermore, the effective interaction potential is long-ranged only along the direction specified by the Raman beams (which can in general be different from the collision axis).

In order to qualitatively understand the size of this range it is useful to consider two limiting situations.  Firstly, atoms scattering with a small momentum difference, $\hbar(k_{3,x}-k_{4,x} ) \ll \sqrt{2m\hbar\Omega_R}$, have nearly the same bare spin composition.  Secondly, particles with a large momentum, $\hbar k_x \gg \sqrt{2m\hbar\Omega_R}$, are unaltered by the Raman coupling.  In either case, $\chi\approx 1$ and the atoms collide as they did absent Raman coupling ({\it 17}).  For intermediate momenta $\hbar k_x\approx\sqrt{2m\hbar\Omega_R}$ \,there is significant dressing-altered scattering, giving a range $\ell_R\approx\hbar/\sqrt{2 m\hbar\Omega_R} = \sqrt{E_L / \hbar\Omega_R k_L^2}$, to the effective interaction potential, with $\ell_R\approx 50\nm\gg\ell_{\rm vdW}$ for our experiment.  This expression is valid in the limit $\hbar\Omega_R\gtrsim 4 E_L$, required for restricting atoms to the lowest band for our collision energies.  Whereas slowly moving atoms experience $s$-wave, scattering higher partial waves are possible at slightly larger velocities.  This is because the range of the effective potential is much larger than that of the real potential, and hence the velocity for which higher partial waves become important is proportionally smaller.

To gain further insight into the nature of the real space representation of the effective interaction potential a quantitative analysis is necessary.  An analytical approximation to the real space screened potential can be found for $\hbar\epsilon = -4 E_L$.  While this is not the case for the experiments described in this paper (where $\hbar\epsilon = 0.42 E_L$), this limiting case is useful for pedagogical reasons.  The form of the real space potential for general $\epsilon$ can be solved numerically and has a similar form to the analytical solution we derive below.

In the following discussion we consider the component of an atom's momentum along the direction of the Raman beams, dropping the subscript $x$, that is $k_x=k$.  For $\hbar\epsilon=-4 E_L$, Hamiltonian \ref{eq:matrixHamiltonian} is
\begin{align}
\check H &= \mathbf{k}^2\check 1 + \left(4 k -\delta\right) \check F_z + \tilde\Omega\check F_x,
\end{align}
where ${\mathbf F} = (\check F_x, \check F_y, \check F_z)$ is the total angular momentum vector, $\tilde\Omega = \Omega/\sqrt{2}$, and we have used dimensionless variables $\Omega\rightarrow \Omega/E_L$, $k\rightarrow k/k_L$.  Owing to the simple form, the problem can be solved using angular momentum algebra and the screened interaction in the lowest dressed state is exactly
\begin{align*}
\chi =& \frac{1}{4}\left\{1 + \frac{1 + 16k_1 k_3/\tilde\Omega^2}{\left[1+16k_1^2/\tilde\Omega^2\right]^{1/2}\left[1+16k_3^2/\tilde\Omega^2\right]^{1/2}}\right\}\nonumber\\
&\times \left\{1 + \frac{1 + 16k_2 k_4/\tilde\Omega^2}{\left[1+16k_2^2/\tilde\Omega^2\right]^{1/2}\left[1+16k_4^2/\tilde\Omega^2\right]^{1/2}}\right\}.
\end{align*}
As is often the case with screening problems, this effective interaction potential is not simply related to a coordinate space potential $V\left(\left|r_1-r_2\right|\right)$.  In the following sections, we will see when such a potential exists and what it is.

Perhaps the most unusual aspect of this potential is that while momentum is still conserved, Galilean invariance is not.  This breakdown of Galilean invariance is common in screening problems where there is a preferred reference frame.  In this case, rapidly moving atoms are Doppler shifted from Raman resonance and the effects of the laser coupling vanish.  

\subsection*{Galilean invariant form}

In our experiment, the dressed atoms are slowly moving, so we expect a Galilean invariant system.  By expanding $\chi$ to second order in the center of mass momentum $k_{\rm com}$, we find
\begin{widetext}
\begin{align}
\chi(k_1,k_2,k_3,k_4) &\approx \frac{1}{4}\left\{1 +
\frac{1 + 4(k_1-k_2) (k_3-k_4)/\tilde\Omega^2}{\left[1+4(k_1-k_2)^2/\tilde\Omega^2\right]^{1/2}\left[1+4(k_3-k_4)^2/\tilde\Omega^2\right]^{1/2}}\right\}^2 + O\left(\frac{k_{\rm com}^2}{\tilde\Omega^2}\right).
\end{align}
\end{widetext}
Because this is only a function of the differences $k_1-k_2$ and $k_3-k_4$, it has recovered Galilean invariance, but only when $(k_{\rm com} E_L/ \kl \Omega)^2$ (expressed in terms of experimental variables) is small.  In our collision experiment, $k_{\rm com} = 0$.

Still, this dielectric function is not of the form $\chi(k_1-k_3)$, so the resulting potential is not just a function of a single distance.  This potential has a fascinating non-local structure in $r$-space which goes beyond the current work.  If we introduce new difference variables, $\delta k_{13} = k_1-k_3$ and $K_{13} = k_1 + k_3$ (and likewise for $k_2$ and $k_4$) and assume that the sum of the initial and final momenta $K_{13}$ is small, we arrive at the expansion
\begin{align}
\chi =& \left[\frac{1}{1+4\left(\delta k_{13}/\tilde\Omega\right)^2}\right]^2 + O\left(\frac{K_{13}^2}{\tilde\Omega^2}\right)\nonumber \\
&+ O\left(\frac{K_{13} K_{24}}{\tilde\Omega^2}\right) + O\left(\frac{K_{24}^2}{\tilde\Omega^2}\right),
\end{align}
expressing the interaction in $k$-space.  This gives rise to the screened potential,
\begin{equation}
\tilde{V}(\delta k) \approx g\left[\frac{1}{1 + 8(\delta k/\Omega)^2} \right]^2.\label{eq:kspacepotential}
\end{equation}
The effective real space potential can be found from the Fourier transform of Eq.~\ref{eq:kspacepotential}, giving
\begin{equation}
V({\mathbf r}) \approx g \frac{\Omega|x|+2\sqrt{2}}{32} \left(\Omega e^{-\Omega|x|/2\sqrt{2}}\right) \delta(y)\delta(z).
\end{equation}
Recall we made $\Omega\rightarrow \Omega/E_L$, $k\rightarrow k/k_L$, and $x\rightarrow x k_L$.  The potential is normalized to $g = 4\pi\hbar^2 a_s/m$ so it correctly represents the $g\delta(x)$ contact interaction as $\Omega\rightarrow\infty$.  The range of the potential is $\approx \lambda/[\pi(\Omega/E_L)]$, which agrees with the qualitative estimate of $\approx 50\nm$ for typical experimental parameters.

\subsection*{Effective pseudopotentials}

In the above, we arrived at an effective potential between atoms.  If instead, we assume all of the momenta are small, but keep the second order terms, we can derive an effective pseudopotential 
\begin{align}
V(x) &\approx g\left[\delta(x) + \frac{16}{\kl^2\Omega^2} \delta''(x)\right]
\end{align}
in which the second-derivative of the delta function leads to $d$-wave scattering.

\end{document}